\documentclass{cccg26}
\usepackage{graphicx,amssymb,amsmath}
\usepackage{multicol, latexsym, amsmath, amssymb}

\usepackage{xcolor}
\usepackage[shortlabels]{enumitem}
\usepackage{hyperref}
\usepackage{subcaption}

\def\defn#1{\textit{\textbf{\boldmath #1}}}

\newcommand{\lemlab}[1]{\label{lemma:#1}}
\newcommand{\lemref}[1]{\ref{lemma:#1}}
\newcommand{\thmref}[1]{\ref{thm:#1}}
\newcommand{\thmlab}[1]{\label{thm:#1}}

\newcommand{\figlab}[1]{\label{fig:#1}}
\newcommand{\figref}[1]{\ref{fig:#1}}

\def\a{{\alpha}}

\def\C{{\mathcal{C}}}

\newcommand{\hide}[1]{}

\newcommand{\JOR}[1]{} 
\newcommand{\Anna}[1]{} 
\newcommand{\erik}[1]{} 
\newcommand{\FF}[1]{} 
\newcommand{\HUGO}[1]{} 
\newcommand{\fade}[1]{} 
\newcommand{\jor}[1]{{\color{black} #1}}
\newcommand{\anna}[1]{{\color{black} #1}}
\newcommand{\ff}[1]{{\color{black} #1}}

\title{Overlapping Unfoldings of Cones and Convex Polyhedra}

\author{%
MIT CompGeom Group\thanks{%
Artificial first author to highlight that the other authors
(in alphabetical order) worked as an equal group.
 }
 \and
 Hugo A. Akitaya\thanks{%
 U. Mass. Lowell, \texttt{hugo\_akitaya@uml.edu}}
 \and
 Erik D. Demaine\thanks{%
 MIT, \texttt{edemaine@mit.edu}}
 \and
 Fabian Frei\thanks{%
 MIT, \texttt{ffrei@mit.edu}}
 \and
 Stefan Langerman\thanks{%
 U. Libre de Bruxelles, \texttt{sl@slef.org}}
 \and
 Anna Lubiw\thanks{%
 U. Waterloo, \texttt{alubiw@uwaterloo.ca}}
 \and
 Joseph O'Rourke\thanks{%
 Smith College, \texttt{jorourke@smith.edu}}
 }

\index{Akitaya, Hugo A.}
\index{Demaine, Erik D.}
\index{Frei, Fabian}
\index{Langerman, Stefan}
\index{Lubiw, Anna}
\index{O'Rourke, Joseph}

\begin{document}
\thispagestyle{empty}
\maketitle

\begin{abstract}
Research on D{\"u}rer's problem focuses on edge unfoldings of
convex polyhedra that avoid overlap.
We invert the goal and find
unfoldings that overlap at some point to any given thickness $t$.

We have two main results.
The first is that,
if we allow unfolding cuts that do not follow polyhedron edges, then there is a 
convex polyhedron that can unfold with overlap of any given thickness.
The second result is that for any given thickness,
there is a convex polyhedron with an edge unfolding that overlaps to that thickness. 
\end{abstract}

\section{Introduction}
%
D{\"u}rer's problem~\cite[Open Problem~21.1]{do-gfalop-07} goes back to a treatise by Albrecht Dürer~\cite{d-pm-1525} from 1525, 
was fully formalized by Shephard~\cite{shephard1975} in 1975, 
and has remained unsolved. It asks whether every convex polyhedron has an
\defn{edge unfolding} (i.e., 
an unfolding where only cuts along edges are allowed) to a \defn{net}, that is, to a simple (i.e., nonoverlapping) 
planar polygon.
\FF{One thing that was confusing me...}\JOR{hiding comment which I think is resolved.}
%

Among the strongest positive results is Ghomi's proof that every convex polyhedron can be stretched affinely into a new polyhedron that has an edge unfolding to a net.
A type of inverse to Ghomi's theorem is the result that every combinatorial convex polyhedron
has a metric realization that has an edge unfolding that overlaps~\cite{AlwaysOverlapJGT}.

\Anna{Let's also introduce general unfoldings hereabouts.  Please rewrite as you like.}
\JOR{Good as it stands.}
\anna{There are convex polyhedra for which every edge unfolding provides a net---for example, the Platonic solids \jor{\cite{hs-eupsn-11}}.  On the other hand, there are polyhedra with overlapping edge unfoldings, for example a skinny tetrahedron~\cite{nf-u3dcp-93} (or see~\cite{do-gfalop-07}), and a recently discovered family of convex polyhedra with regular faces~\cite{shiota2024overlapping}.  In these examples at most two layers overlap at any point.

In a \defn{general unfolding} the cuts are not restricted to edges of the polyhedron; the unfolded surface
\Anna{we could add: where every dihedral angle is opened to $\pi$}\JOR{I think it's clearer w/o mentioning dihedral angle.} is still required to be flat and connected; equivalently, the cuts form a tree that includes all the vertices of the polyhedron.  Every convex polyhedron has a non-overlapping general unfolding, specifically,
the source or star unfolding~\cite{do-gfalop-07}.}

In this paper, we examine whether overlaps of larger thickness
are possible
\anna{for \jor{edge or for} general unfoldings} and, if so, what the limits on overlap
thickness are. 
The first question is answered positively by 
the unfolding of a convex polyhedron in 
Fig.~\figref{TripleOverlap}.
Before addressing the second question, we define the concept of overlap thickness more formally. 

Given an unfolding of a polyhedron 
to the Euclidean plane, define this unfolding's
\defn{thickness} 
of overlap at a point of the plane as the number of distinct, non-boundary points of the surface unfolded to this point (i.e., the number of layers or plies 
of the unfolding at this point).
More precisely, thickness $t$ at a point $p$ means that a line orthogonal to the plane
pierces the interior of the unfolded surface $t$ times.
\Anna{I don't think this sentence is needed, ...}\JOR{Agree, removing sentence.}
Thickness $0$ at a point means that the line misses the interior of the unfolding.
The thickness of an unfolding is its maximum thickness across all points. 
An unfolding of thickness~$1$ has no overlap, the goal of D{\"u}rer's problem for any given polyhedron. 
 
\begin{figure}[htbp]
\centering
\includegraphics[width=0.9\columnwidth]{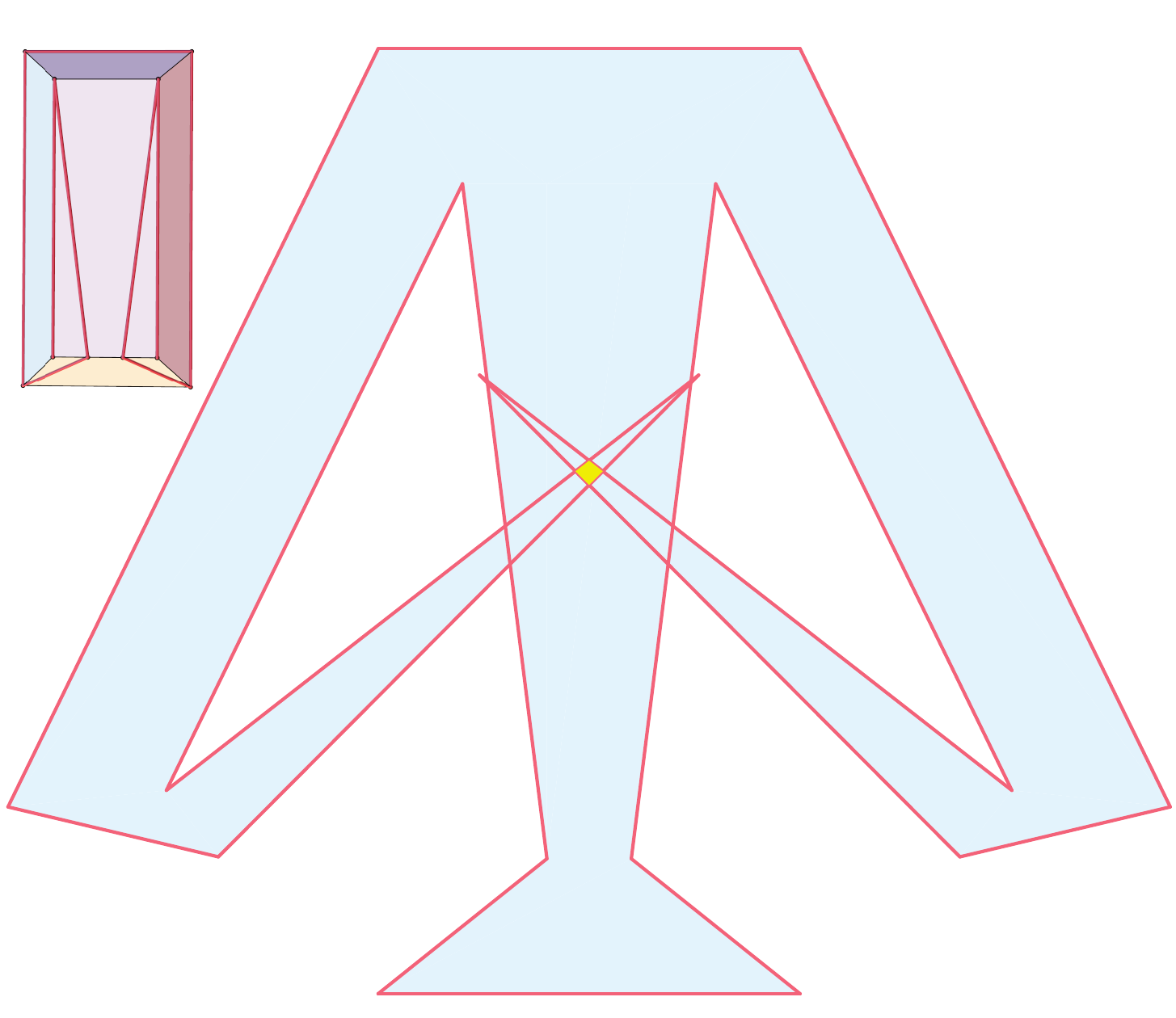}
\caption{A rectangular prismoid shown from the top with the red lines showing a cut tree and the resulting unfolding with triple overlap in the yellow diamond. \ff{The bottom face is omitted in the picture.}} 
\figlab{TripleOverlap}
\end{figure}

\medskip
We have two main results: The first concerns general unfoldings of a convex polyhedron,
and the second concerns edge unfoldings of convex polyhedra.
Both are established through explicit geometric constructions.


\begin{theorem}
\thmlab{ConeOverlap}
There is a convex polyhedron $P$
such that for any $t \in \mathbb{N}$, $P$~can be unfolded using general cuts
(so not restricted to edges)
to have overlap thickness $t$.
\end{theorem}

\noindent
Theorem~\thmref{ConeOverlap} will follow from
Lemma~\lemref{ConeIrrational},
which builds on Lemma~\lemref{ConeRational}.
\ff{Both lemmas concern an unbounded cone $\C$ without polyhedral structure, i.e., just a single cone vertex.}

\begin{lemma}
\lemlab{ConeRational}
For a cone $\C$ of apex curvature\footnote{
The \defn{curvature} of a vertex is its ``angle gap'':
$2\pi$ minus the incident surface angle.}
$\omega= \lambda \cdot 2\pi$,
with $\lambda=a/b$
in lowest terms, $a,b \in \mathbb{N}$,
there is a cut path that unfolds the cone to have overlap thickness
$t=b-a$, and this is the maximum possible.


\hide{
There is a 
cut path 
unfolding $\C$ to have overlap thickness
$t=b-a$, and this is the maximum possible.}
\end{lemma}

\begin{lemma}
\lemlab{ConeIrrational}
For a cone $\C$ whose apex curvature is an irrational multiple of $\pi$,
there is, for any 
$t \in \mathbb{N}$, a 
cut path 
\anna{that unfolds}
$\C$ to have overlap thickness $t$.
\end{lemma}
\noindent

%
%
\jor{
These cone unfoldings are not edge unfoldings.
}
\FF{This sentence suggests to me that the unfolding of Theorem 1 might be an edge unfolding.}
\jor{
Our second main result 
concerns
edge unfoldings:
}

\begin{theorem}
\thmlab{PolyhedronOverlap}\label{PolyhedronOverlap}
For any 
$t\in \mathbb{N}$, 
there is a convex polyhedron $P$ of $n=O(t)$ vertices
with an edge unfolding achieving overlap thickness of at least $t$.
\end{theorem}

\Anna{a more precise version:}
\anna{The bound in the theorem is tight: to achieve overlap thickness $t$ with an edge unfolding the polyhedron must have at least $t$ faces, and thus $\Omega(t)$ vertices.}
\Anna{My intention was that the previous sentence would replace the following:}
\fade{Note that the overlap thickness of an edge unfolding is at most the number of faces. 
\ff{Since the numbers of vertices, edges, and faces of convex polyhedra 
\Anna{differ by a constant factor},
linearly bound each other,\JOR{I prefer this linearly bound phrasing.} Theorem~\ref{PolyhedronOverlap} is essentially tight.} 
}

\section{Cone Overlap}
\bigskip
For intuition, we
first describe a construction establishing \jor{Lemma~\lemref{ConeRational}} 
when $\omega = \pi/2$, 
i.e., $a=1,b=4$,
achieving triple overlap, $t=3$.
The cone $\C$ can then be viewed as a corner of a cube, 
although no polyhedral structure is relevant 
except to suggest labeling.

\begin{figure}[htbp]
\centering
\includegraphics[width=0.8\columnwidth]{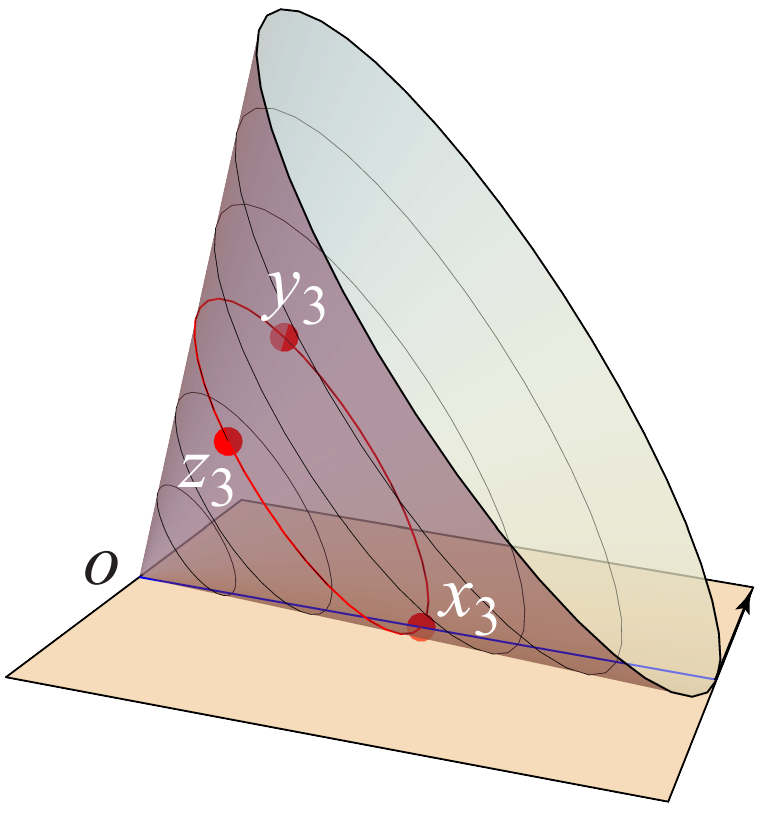}
\caption{Rolling the cone on the plane about the apex at the origin $o$ twice lets each point of $z_3,y_3,x_3$ 
(all on the red circle)
touch the same point in the plane once.
}
\figlab{ConeRoll}
\end{figure}

Imagine laying $\C$ on the plane so that the apex is at the origin $o$,
as illustrated in Fig.~\figref{ConeRoll}. 
In any unfolding of $\C$, a point on $\C$ at distance $d$ from the apex maps to a point in the plane at distance $d$ from the origin.
In Fig.~\figref{ConeRoll}, points on the red circle map to points in the plane on a circle obtained by rolling $\C$ about the origin. We first show that the three equally spaced points $x_3, y_3, z_3$
can map to the same point on the plane, and then show that there is an unfolding that achieves this.
Suppose we roll the cone 
so that its contact with the plane rotates counterclockwise
(i.e., 
away from the viewer, in the direction of the arrow in Fig.~\figref{ConeRoll}), beginning with $x_3$ on the plane.  After rolling by $2\pi$,
point $y_3$ lands on top of the initial $x_3$---rolling $3\pi/2$ is one complete
roll of $\C$, and a further $\pi/2$ consumes the separation between $x_3$ and $y_3$ around the red circle.
Another roll of 
$2\pi$ brings $z_3$ to land on 
the original $x_3$,
achieving triple overlap at that point.
\anna{Looking ahead, the generalization for Lemma~\lemref{ConeRational} will be that $t=b-a$ equally spaced points around a cone of apex curvature $(a/b)\cdot 2\pi$ can develop by rolling so that they overlap.}

We now show that there is an unfolding of the cone
that achieves this triple overlap.  We cut a non-self-crossing path on the cone that goes through $x_3, y_3, z_3$ so that the three points coincide in the resulting unfolding.
The path begins at $x_3$ and spirals outward and clockwise as shown in 3D in Fig.~\figref{Spiral90_CutPath_3D} and in 2D (with the cone sliced along $o x_3$) in Fig.~\figref{Spiral90_3paths}.  
To describe a specific polygonal spiral, we place circles $C_i, i=1,\ldots,6$ at regular intervals centered at $o$, with $x_3,y_3,z_3$ on $C_3$, and we place the bends of the path at the intersections of the circles with the three rays from $o$.
The spiral path begins at $x_3$, intersects the rays $o y_3$ and $o z_3$ inside $C_3$, then crosses the ray $o x_3$ outside $C_3$. 
After that 
the path
goes to $y_3$, then follows just outside the previous part of the spiral until reaching $z_3$.
To unfold the cone, we must extend the path to reach $o$---we use segment $x_3 o$---and to reach the boundary---we use a segment from $z_3$ to the boundary. 
Cutting the path splits $x_3, y_3$ and $z_3$ into two points each, and both the inner and outer (relative to the spiral) three copies coincide in the unfolding. 
Since these are boundary points of the unfolding, they do not count as thickness 3; however, the slightly displaced cyan points shown in Fig.~\figref{Spiral90_3paths} give a point of overlap thickness 3 in the unfolding.
\begin{figure}[htbp]
\centering
\includegraphics[height=1.0\columnwidth]{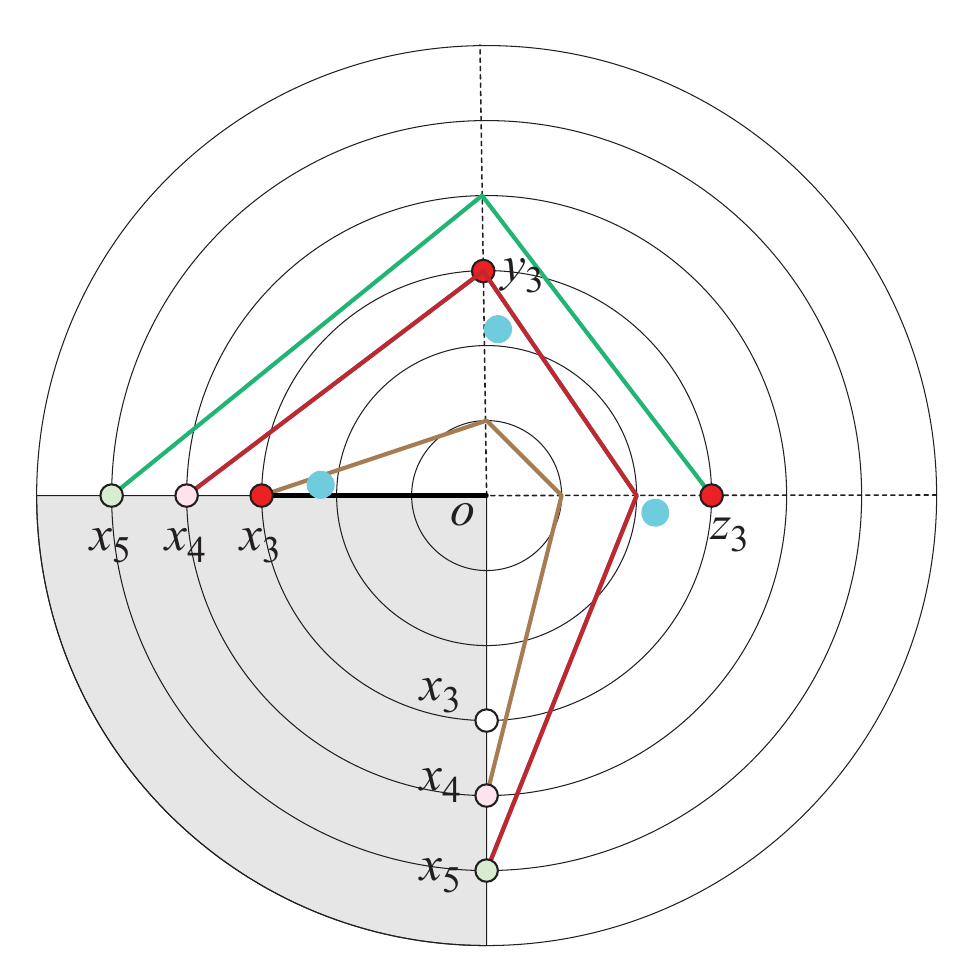}
\caption{
\jor{Cone flattened. Grey quadrant reflects the angle gap of $\pi/2$ at $o$.
Cyan disks will overlap when path is cut and unfolded.}
}
\figlab{Spiral90_3paths}
\end{figure}

\hide{A view of the cone in 3D is shown in Fig.~\figref{Spiral90_CutPath_3D}.
The three marked points $z_3, y_3, x_3$ will lie on top of one another when unfolded.}

\begin{figure}[htbp]
\centering
\includegraphics[width=0.90\columnwidth]{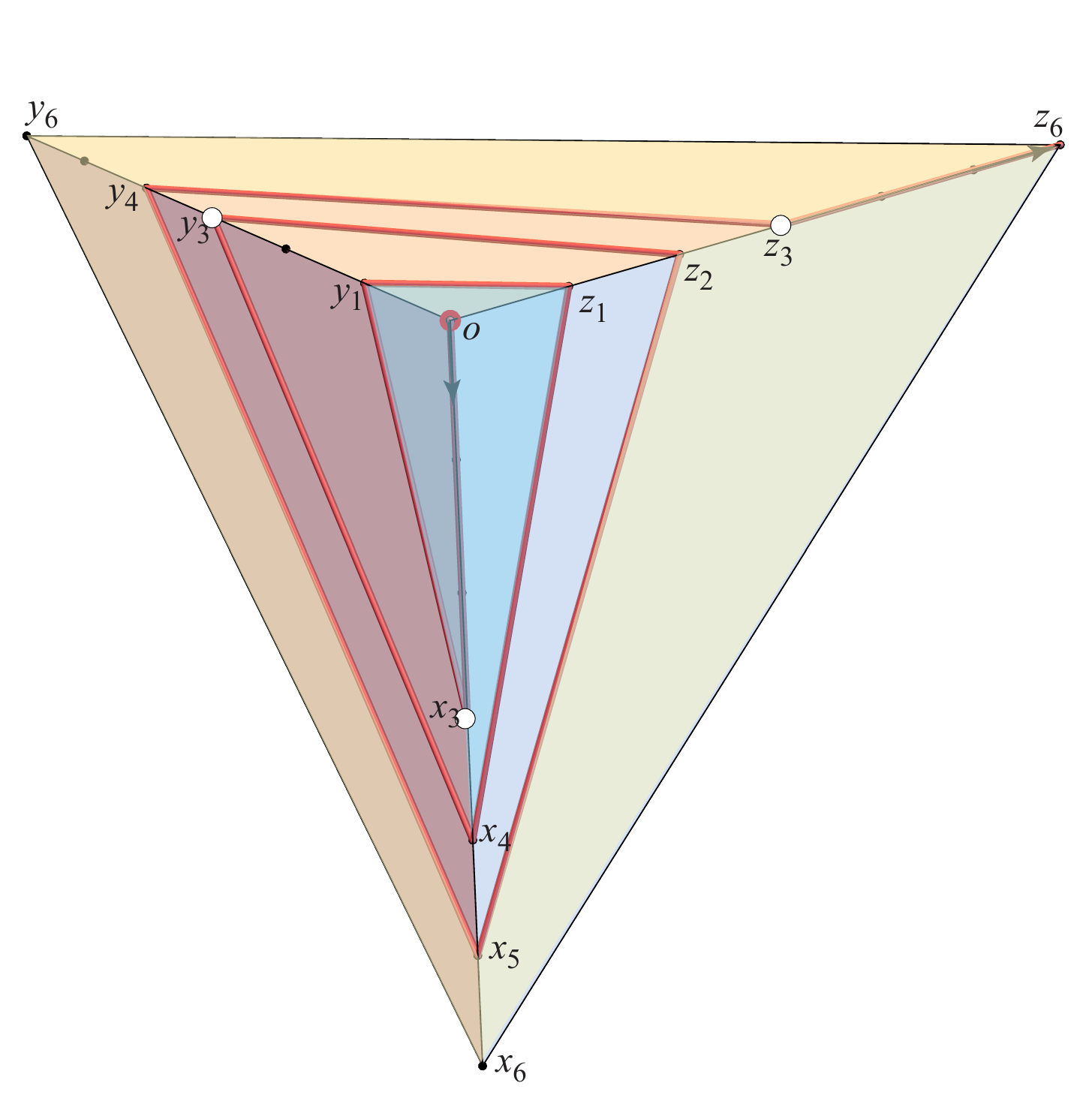}
\caption{Cut path: 
$(o, x_3, y_1,\ldots, z_3, z_6)$.
Marked points
$x_3,y_3,z_3$ coincide in the unfolding.
}
\figlab{Spiral90_CutPath_3D}
\end{figure}

\begin{figure}[htbp]
\centering
\includegraphics[width=1.0\columnwidth]{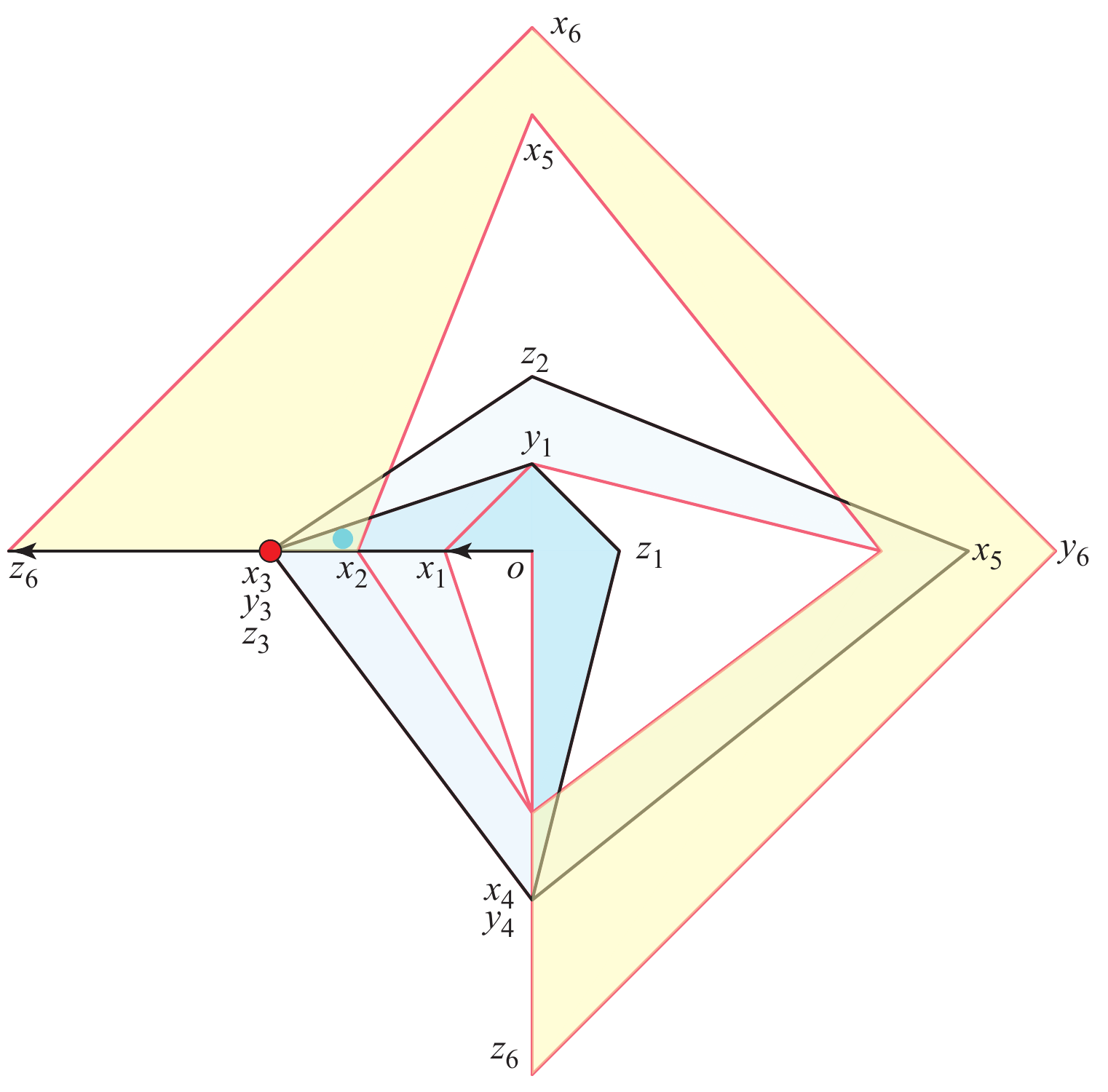}
\caption{Cutting the path in Fig.~\figref{Spiral90_CutPath_3D} 
results in triple overlap at 
\jor{the cyan point near} $x_3=y_3=z_3$.
}
\figlab{Spiral90_Layout}
\end{figure}


\Anna{I suggest moving Figure 8 before Figure 7....}\JOR{Will do.}

\begin{figure}[htbp]
\centering
\includegraphics[width=1.0\columnwidth]{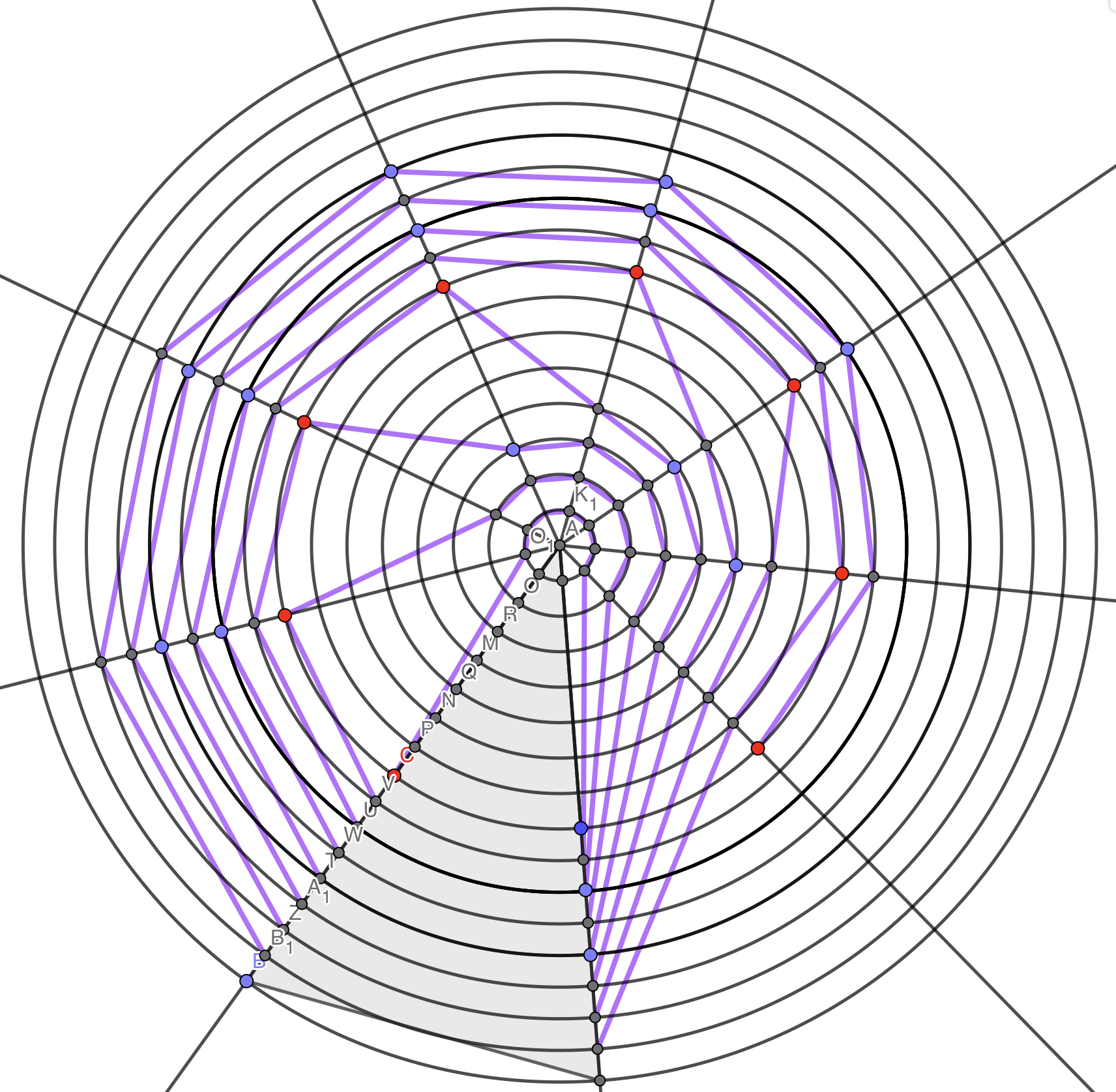}
\caption{$\omega=2\pi/9$ achieves overlap thickness $t=8$.}
\figlab{Stefan9SpiderWeb}
\end{figure}

Fig.~\figref{Stefan9SpiderWeb} illustrates the analog\ff{ue} of Fig.~\figref{Spiral90_3paths} 
for the case $a=1, b=9$, so $\omega= 2\pi/9 = 40^\circ$, where we achieve overlap thickness~$8$.
The $8$ red points 
lie on a circle of radius $8$ 
spaced at intervals of $2\pi/9$.  The cut path starts at the origin and spirals outward, passing through the $8$ points in clockwise order and ending at the cone boundary.


\begin{figure}[htbp]
\centering
\begin{subfigure}{0.5\columnwidth}
\includegraphics[width=1.0\columnwidth]{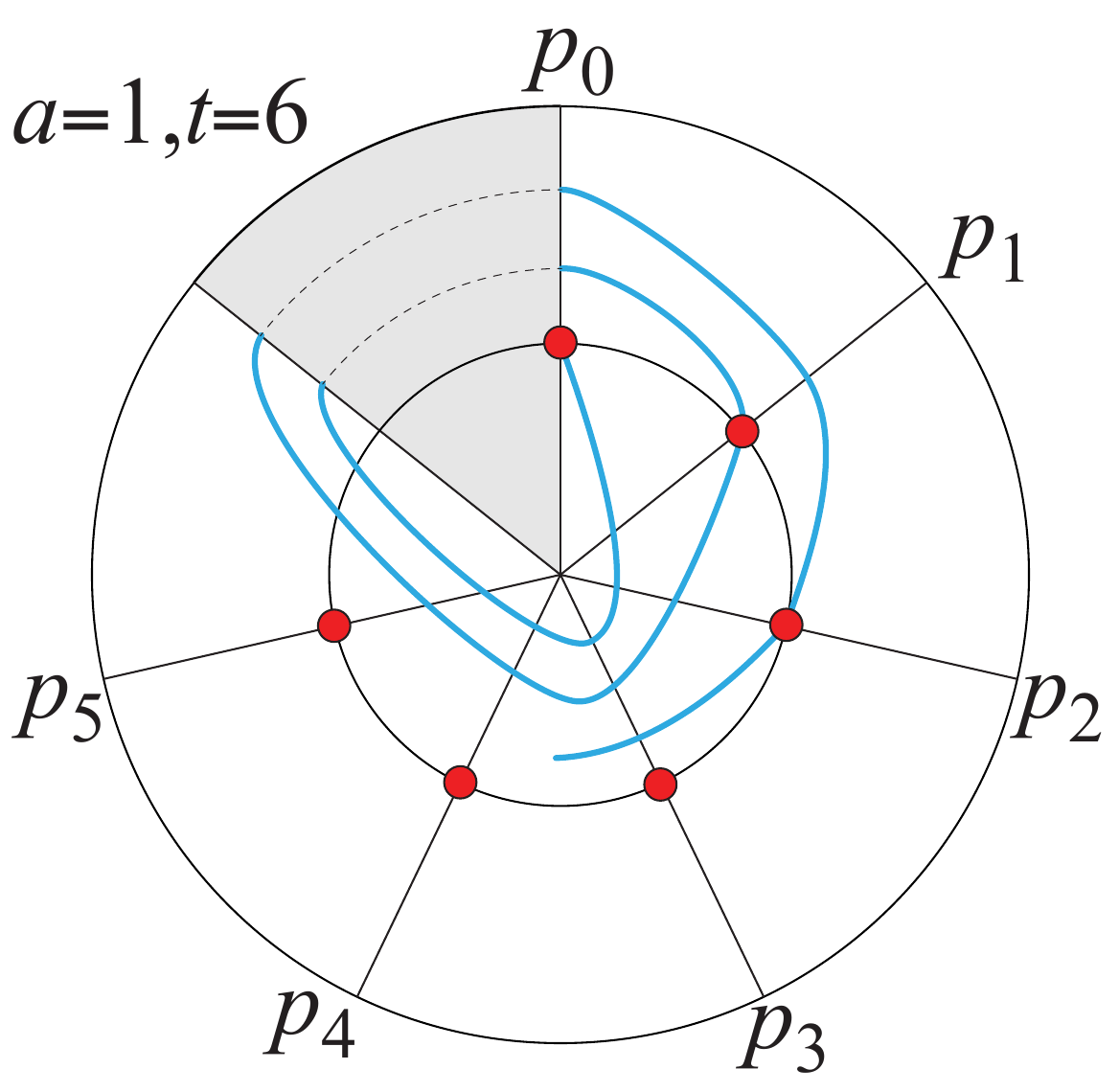}
%
\begin{minipage}{1.0\columnwidth}
\begin{center}
\begin{align*} 
q_1 &= 7 \bmod 6 = 1 \\ 
q_2 &= 14 \bmod 6 = 2 \\
q_3 &= 21 \bmod 6 = 3 \\
q_4 &= 28 \bmod 6 = 4 \\
q_5 &= 35 \bmod 6 = 5 
\end{align*}
\end{center}
\end{minipage}
\end{subfigure}%
\begin{subfigure}{0.5\columnwidth}
\centering
\includegraphics[width=1.0\columnwidth]{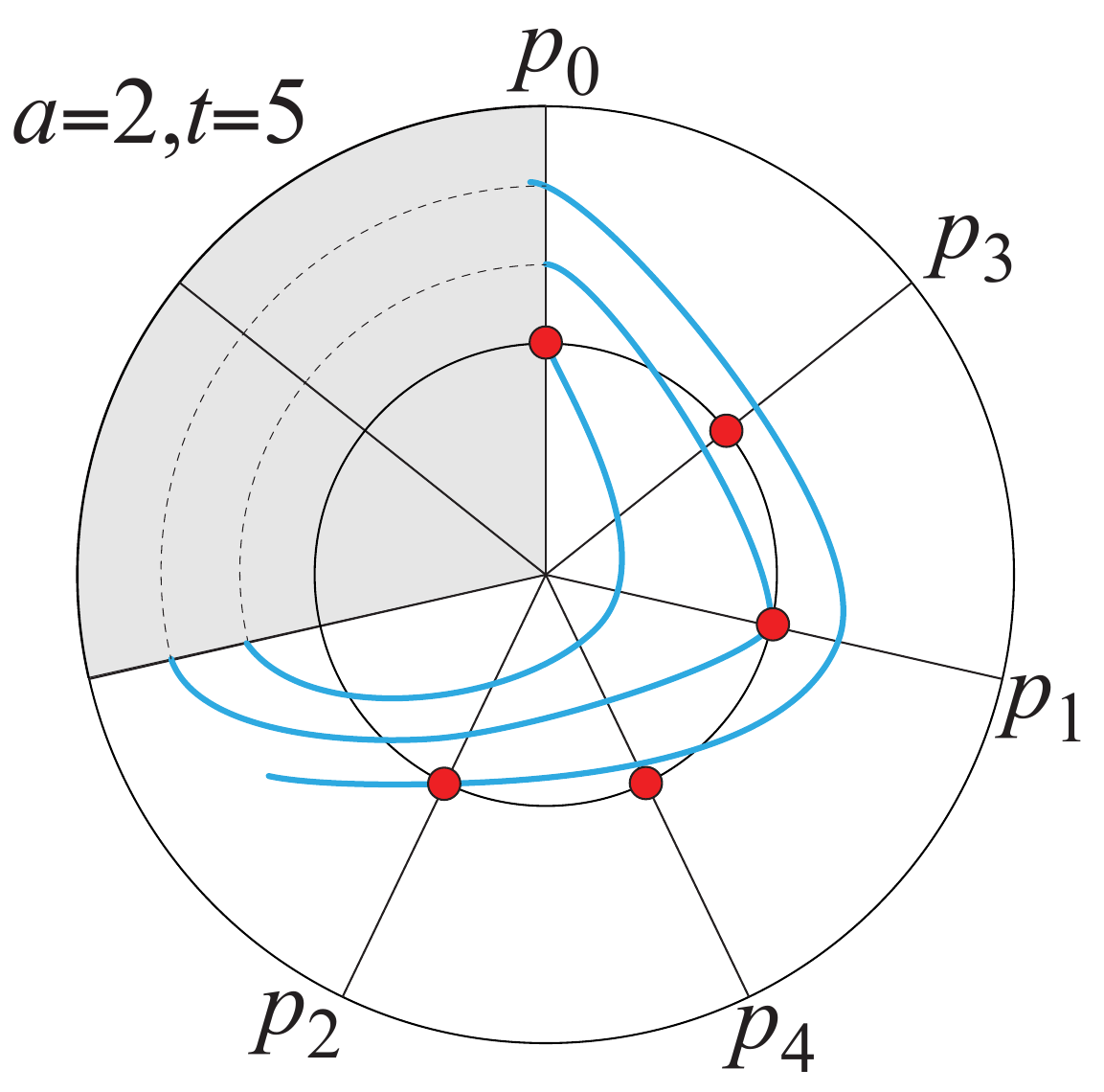}
%
\begin{minipage}{1.0\columnwidth}
\begin{center}
\begin{align*} 
q_1 &= 7 \bmod 5 = 2 \\ 
q_2 &= 14 \bmod 5 = 4 \\
q_3 &= 21 \bmod 5 = 1 \\
q_4 &= 28 \bmod 5 = 3 \\
\mbox{} &
\end{align*}
\end{center}
\end{minipage}
\end{subfigure}
\end{figure}
%
%
\addtocounter{figure}{-1} 
\FF{Is there a reason not to include both parts in one figure? To allow them to be in separate columns? I guess then it would still be a bit strange to have only one caption in the new column.}
\begin{figure}[htbp]
\centering
\begin{subfigure}{0.5\columnwidth}
\includegraphics[width=1.0\columnwidth]{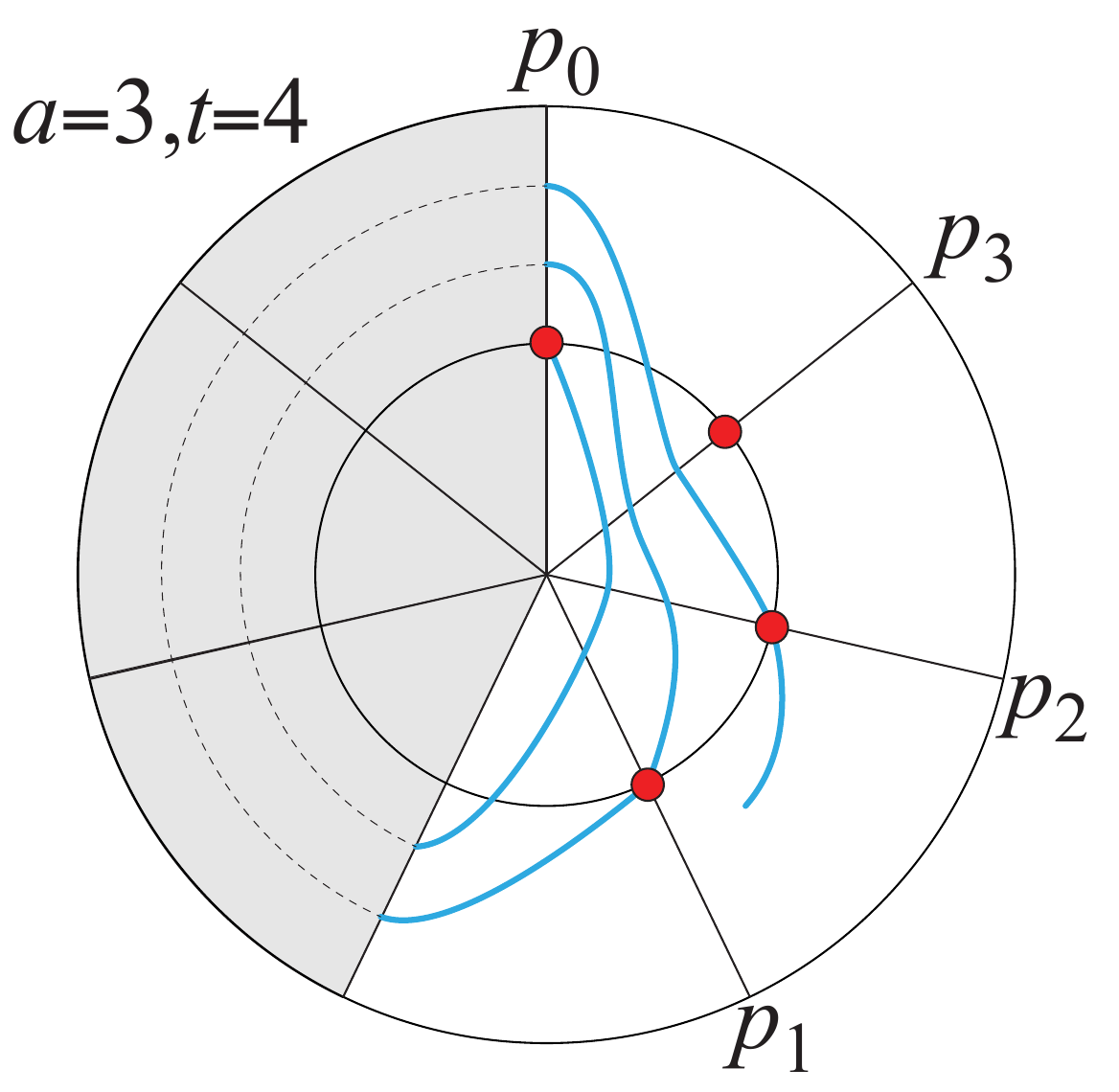}
%
\begin{minipage}{1.0\columnwidth}
\begin{center}
\begin{align*} 
q_1 &= 7 \bmod 4 = 3 \\ 
q_2 &= 14 \bmod 4 = 2 \\
q_3 &= 21 \bmod 4 = 1 
\end{align*}
\end{center}
\end{minipage}
\end{subfigure}%
\begin{subfigure}{0.5\columnwidth}
\centering
\includegraphics[width=1.0\columnwidth]{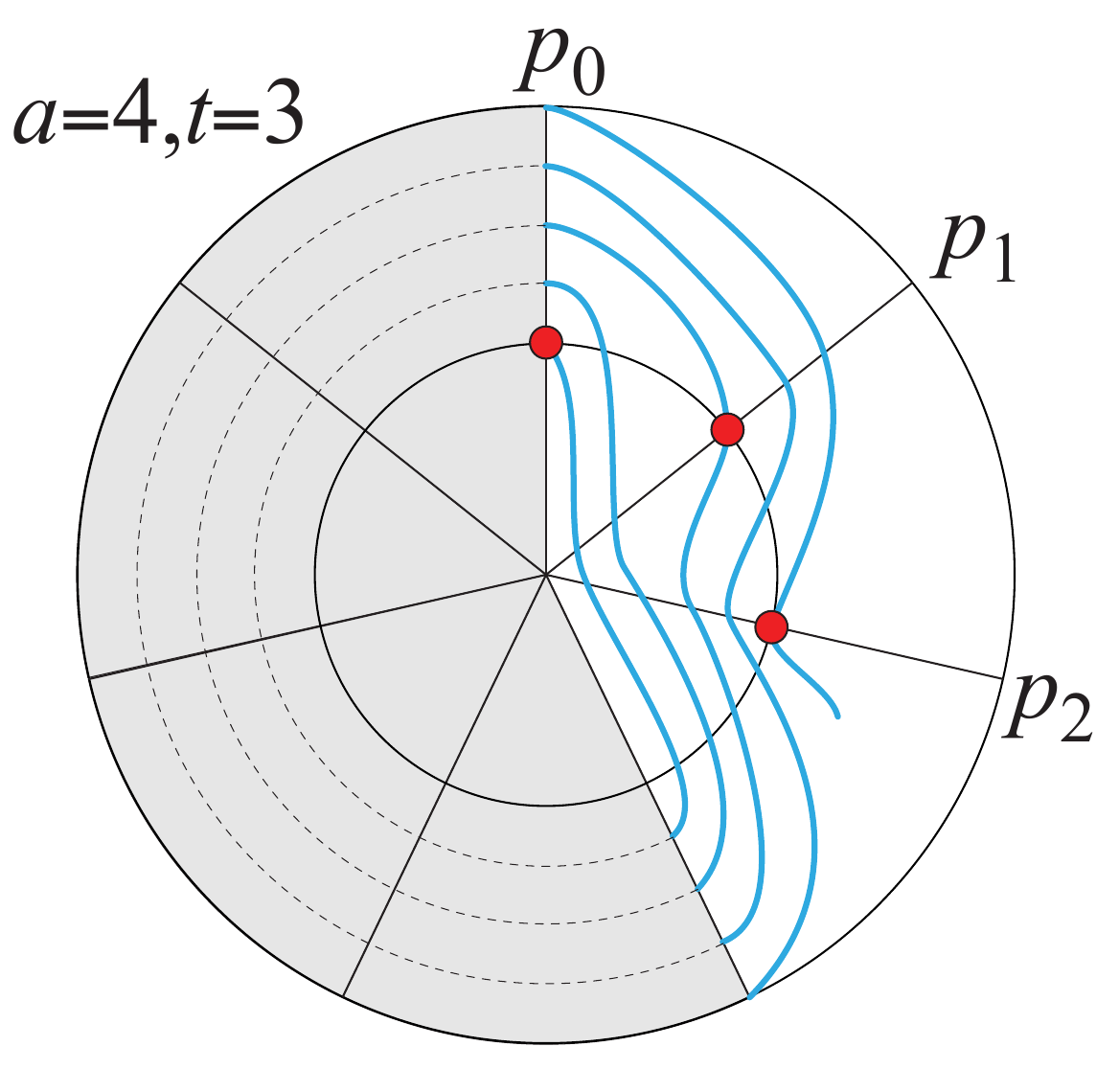}
%
\begin{minipage}{1.0\columnwidth}
\begin{center}
\begin{align*} 
q_1 &= 7 \bmod 3 = 1 \\ 
q_2 &= 14 \bmod 3 = 2 \\
\mbox{} &
\mbox{} &
\end{align*}
\end{center}
\end{minipage}
\end{subfigure}
\caption{Examples for $b=7$ and $a=1,2,3,4$, \anna{showing the points of overlap in red and the beginning of the cut path in cyan.} 
The points of overlap are 
$p_i = (b\,i \bmod(b-a)) \cdot 2\pi /b$.  For brevity we list $q_i = b\,i \bmod(b-a)$.}
\figlab{Spiral7}
\end{figure}

\Anna{new:}
We now turn to the general case of Lemma~\lemref{ConeRational}, 
where $\lambda = a/b$.
See Fig.~\figref{Spiral7}.  The circle is divided by $b$ spokes into $b$ sectors each of angle $2\pi/b$.  The cone occupies $b-a$ sectors. 
We will find a cut path that makes $b-a$ points overlap in the unfolding.  The points lie on a circle centered at the origin, one point on each spoke on the cone.  When $a=1$, the cut path goes through the $b-1$ points in cyclic order---by the angle measured clockwise from $p_0$ these are $p_0=0$, $p_1 = \frac{1}{b}\cdot 2\pi, p_2=\frac{2}{b}\cdot 2\pi, \ldots, p_{b-2} = \frac{b-2}{b}\cdot 2\pi$.  More generally, when $a \ne 1$, 
the order may be different.  
As we roll the cone, the first point that will overlap with $p_0 =0$ is $p_1 = (1 \bmod{\frac{b-a}{b}})\cdot 2\pi$. 
The $i$th point is $p_i = (i \bmod{\frac{b-a}{b}})\cdot 2\pi$ or, equivalently, $(b\,i \bmod{(b-a)}) \cdot 2\pi /b$.
The cut path spirals outward, visiting the points $p_i$ in order.  The number of times the path spirals around the cone between $p_i$ and $p_{i+1}$ is $\lfloor b/(b-a) \rfloor$; for example, $a=4,b=7$ requires
\ff{spiraling around the cone twice} 
as shown in Fig.~\figref{Spiral7}. 

Finally, we note that for $\lambda = a/b$ it is not possible to have 
\anna{overlap thickness greater than $b-a$}
because 
\anna{$p_{b-a} = p_0$.}

\Anna{new:}
We now turn to Lemma~\lemref{ConeIrrational}.
When $\C$ is a cone whose apex curvature $\omega= \lambda 2\pi$ is an irrational multiple of $\pi$, we will create a
cut path that produces an unfolding with overlap thickness $t$ for any given $t \in \mathbb{N}$.
As in the above case where $\lambda = a/b$, we identify $t$ points starting with the initial point $p_0 = 0$.
\jor{The angle of the surface of the cone is $(1-\lambda) 2\pi$, and we seek the remainder after division by this expression.
So when} we roll the cone, the first overlap with $p_0$ is at $p_1= (1 \bmod{(1-\lambda)})\cdot 2\pi$.
The $i$th point is $p_i= (i \bmod{(1-\lambda)})\cdot 2\pi$.
Unlike for $\lambda = a/b$, this sequence never repeats, so we can continue to $p_{t-1}$ for any $t$.
As above, we can make a cut path that spirals outward and visits $p_0, \ldots, p_{t-1}$ in order.

\medskip
%

Theorem~\thmref{ConeOverlap} follows by creating a 
\anna{tetrahedron} whose
apex has irrational curvature \anna{and extending the cut path to cut all but one of the edges of the base.}


\section{Edge Unfolding Overlap}
%

The construction in Fig.~\figref{TripleOverlap}, which brings the tips of thin
triangles to triple overlap 
in the center of the unfolding, can be extended to achieve larger overlap thickness.
Fig.~\figref{Shutter_x4_Overlap} shows a ``camera shutter'' structure that
results in overlap of thickness $t=5$.

It is plausible that generalizing this example to a $k$-sided shutter will achieve
overlap thickness $t=k+1$. 
However, there are two obstacles to proving
Theorem~\thmref{PolyhedronOverlap} via this strategy.
First, controlling the precise curvatures at the vertices that are leaves of
the cut tree, needed to produce overlap in the center of the shutter, is quite delicate.
Second, 
the unfoldings are not edge unfoldings because some vertices have zero curvature
and some edges have flat dihedral angles.
It seems difficult to make all vertices ``real'' (positive curvature) and all cut edge
dihedrals $< \pi$. 
\jor{Indeed, the triangulation graph of the top square might not be ``regular''~\cite[p.55]{drs-tasa-10}, which would imply it cannot be lifted to a convex dome.}
Avoiding these complications, we prove Theorem~\thmref{PolyhedronOverlap} by nesting overlap regions.

\begin{figure}[htbp]
\centering
\includegraphics[width=1.0\columnwidth]{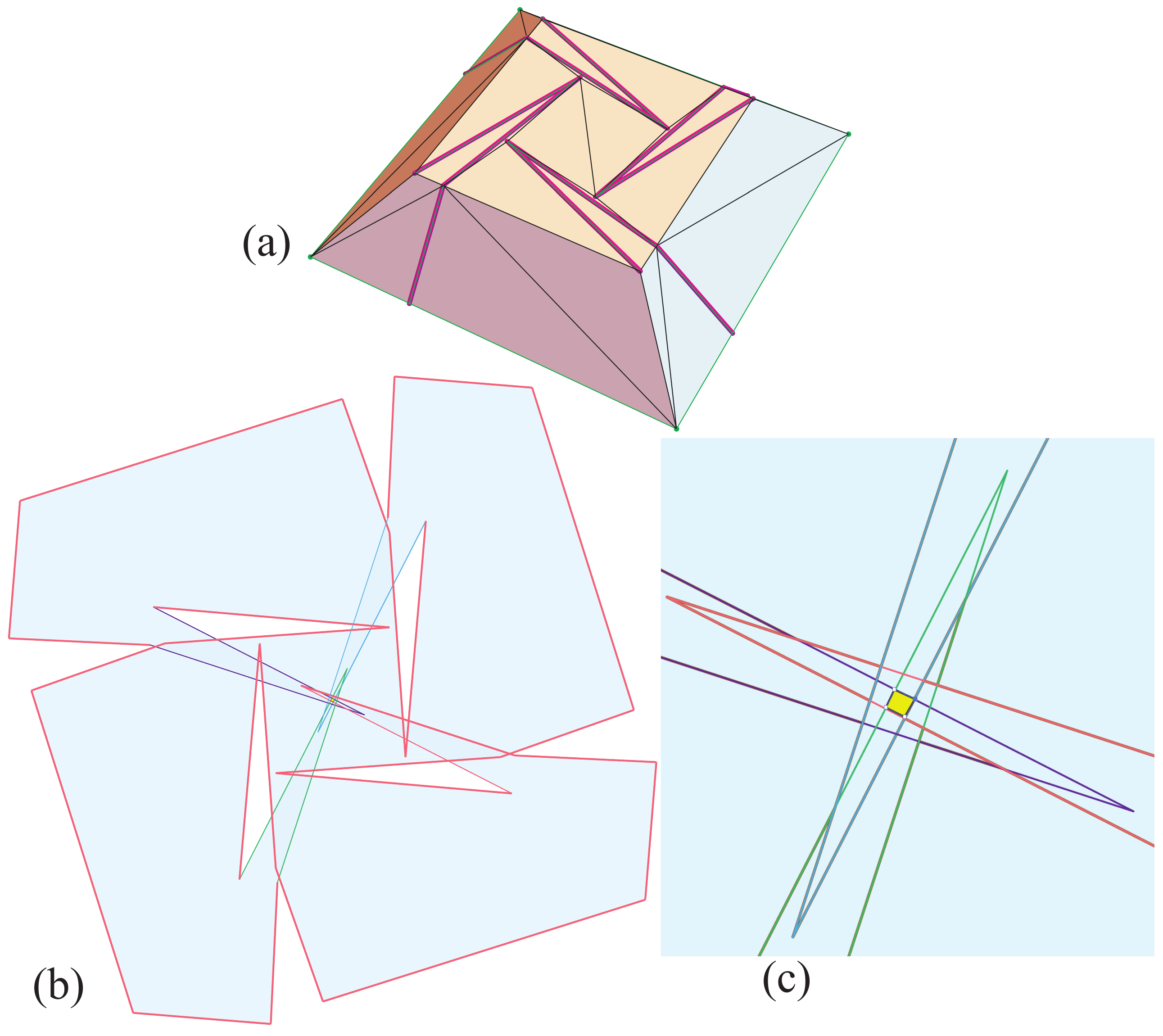}
\caption{(a)~``Shutter'' polyhedron. Black edges are triangulation edges with flat dihedral angles.
(b)~Unfolding when red edges are cut. (The unfolding of the square base is not shown.)
(c)~Yellow square has overlap thickness $t=5$.}
\figlab{Shutter_x4_Overlap}
\end{figure}

\subsection{Overlap Gadget}
Our main tool is an \defn{overlap gadget}, illustrated in Fig.~\figref{Gadget_1st_2D3D}.
It consists of the convex hull of two parallel nested triangles,
$\triangle a_1 b_1 c_1=\triangle_1$ above $\triangle_0$.

There are two constraints.
First, 
$\a = \angle c_1 a_1 b_1$
is acute. The details are easiest to see if $\a$ is not only acute, but small.
The figures use $\arctan(1/2) \approx 26.6^\circ$.
Second, the curvature $\omega$ at vertex $c_1$ is small.
That curvature is controlled by the height \anna{in the $z$ direction}
of $\triangle_1$ over $\triangle_0 $.
\begin{figure}[!h]
\centering
\includegraphics[width=0.8\columnwidth]{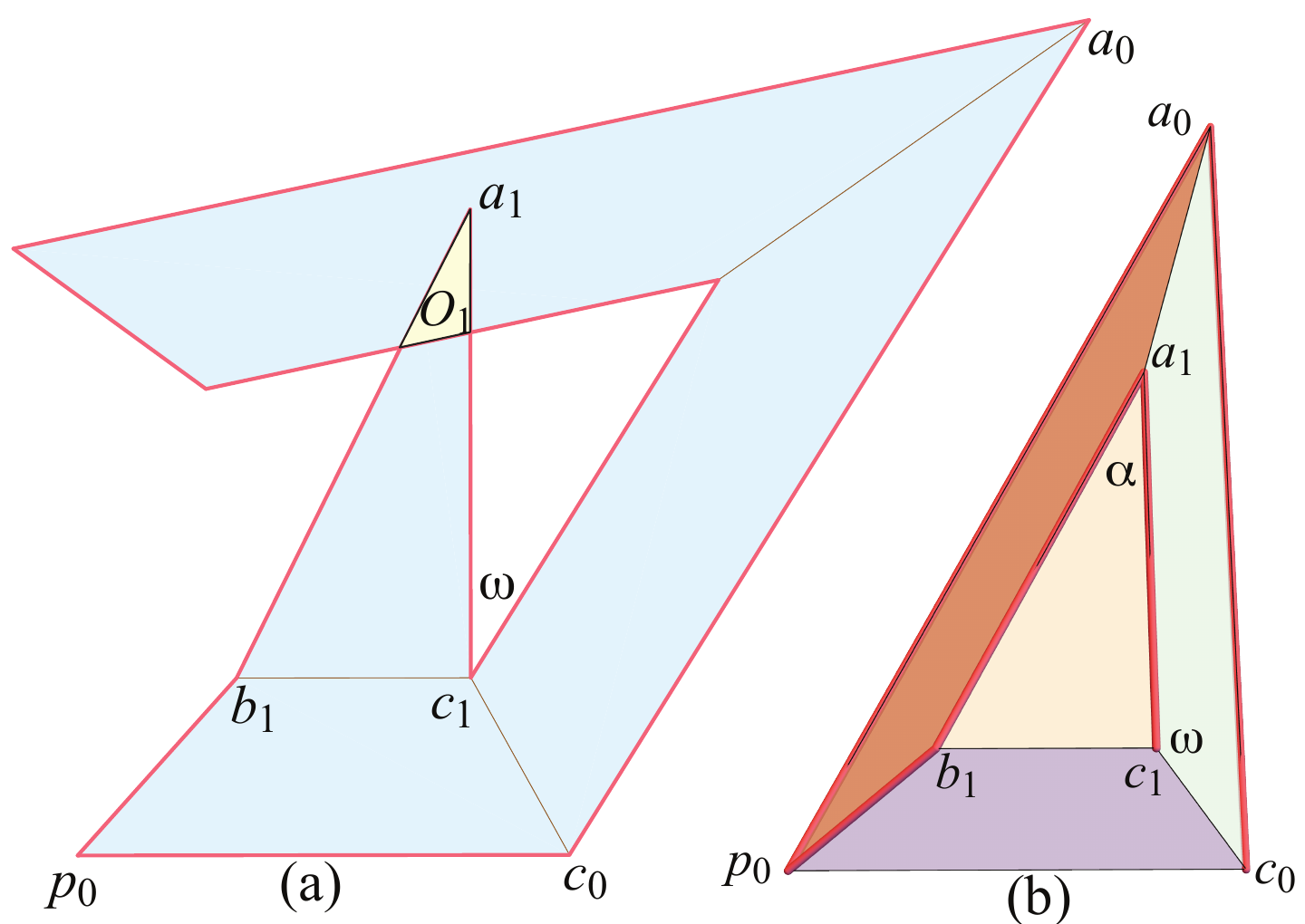}
\caption{(b)~Overlap gadget. Cut path red.
(a)~Overlap $O_1$ in yellow.}
\figlab{Gadget_1st_2D3D}
\end{figure}

The cut tree is a path.
The base triangle folds out below edge $p_0 c_0$.
The unfolding is shown in Fig.~\figref{Gadget_1st_2D3D}(a).
The tip of 
$\triangle_1$ overlaps within the overlap region $O_1$.
We call this 
\ff{first-level} 
gadget $P_1$.
\ff{It achieves }
\ff{an overlap of thickness $t=2$}
near the triangle tip $a_1$.

\subsection{Nesting Gadgets}

Now the plan is to nest another overlap gadget within the overlap region $O_1$.
Each overlap gadget 
$\triangle a_i b_i c_i = \triangle_i$
will create a region of overlap into which a smaller overlap gadget can be placed.

We illustrate achieving $t=3$, building on the $t=2$ overlap in Fig.~\figref{Gadget_1st_2D3D}(a).
Fig.~\figref{Gadget_2nd_3D} shows $P_2$, a modification of $P_1$ in Fig.~\figref{Gadget_1st_2D3D}(b).

Throughout, subscripts indicate the $z$-level of points and regions.
\Anna{suggestion:}
First we add three new vertices, $a_2,b_2,c_2$ forming a triangle $\triangle_2$ that lies slightly above and parallel to $\triangle_1$.\Anna{Are the sides of $\triangle_2$ parallel to the sides of $\triangle_1$? Or does it not matter?}\JOR{In my construction, they were parallel. It certainly makes it simpler. Not certain it's necessary.}\Anna{They are not parallel in Fig.~\figref{Gadget_2nd_3D}(b), see $b_2 c_2$.}\JOR{Right; fixed.}
\jor{The two sides incident to $a_2$ of $\triangle_2$ are chosen parallel to the corresponding sides of $\triangle_1$; see Fig.~\figref{Gadget_2nd_3D}(b).}
We add edges $a_2 a_1, b_2 b_1, c_2 c_1$. The plan is to cut 
from $\triangle_1$ to $\triangle_2$ in the first overlap region, so we add vertex $p_1$ and edge $p_1 b_2$.
Four new vertices are added, $p_1, a_2, b_2, c_2$.
$\triangle a_2 b_2 c_2=\triangle_2$ lies slightly above 
and parallel to $\triangle_1$.
The added vertex $p_1$ connects $\triangle_1$ to $\triangle_2$.
Note the cut tree in Fig.~\figref{Gadget_2nd_3D}(b) is no longer a path:
the branch $p_1, b_2,a_2,c_2$ is attached at $p_1$.

\anna{If $p_1$ were to lie directly on $a_1 b_1$, then edge $p_1 b_2$ would be flat.  Instead, we move $p_1$ slightly to the left of the line $a_1 b_1$ (but still in the plane of $\triangle_1$).}
\anna{Finally, since the perturbation of $p_1$ destroys coplanarity of the quadrilateral $a_1 p_1 b_2 a_2$, we add the edge $p_1 a_2$.  This completes the description of $P_2$.}
\Anna{Or describe complete triangulation of the other quadrilaterals.}\JOR{As you have it is great, esp. adding $p_1 a_2$. Old text faded below.}
\fade{
If $p_1$ were to lie directly on $a_1 b_1$, then cutting from $p_1$ to $b_2$ would not 
follow an edge of the polyhedron $P_2$, because the curvature at $p_1$ is zero.
So $p_1$ is pushed 
out \jor{horizontally} slightly 
in the plane of $\triangle_1$.
\JOR{$p_1$ lies in the plane of $\triangle_1$, pushed horizontally.}
\Anna{I don't understand the next sentence:}
\JOR{You're right, wrong fig ref}
\jor{In Fig.~\figref{Gadget_2nd_2D_all}}, the slight bump-out is discernible.
This change to $p_1$'s location can be arbitrarily small, and is barely visible in the
figures.
}

\begin{figure}[!h]
\centering
\includegraphics[width=0.8\columnwidth]{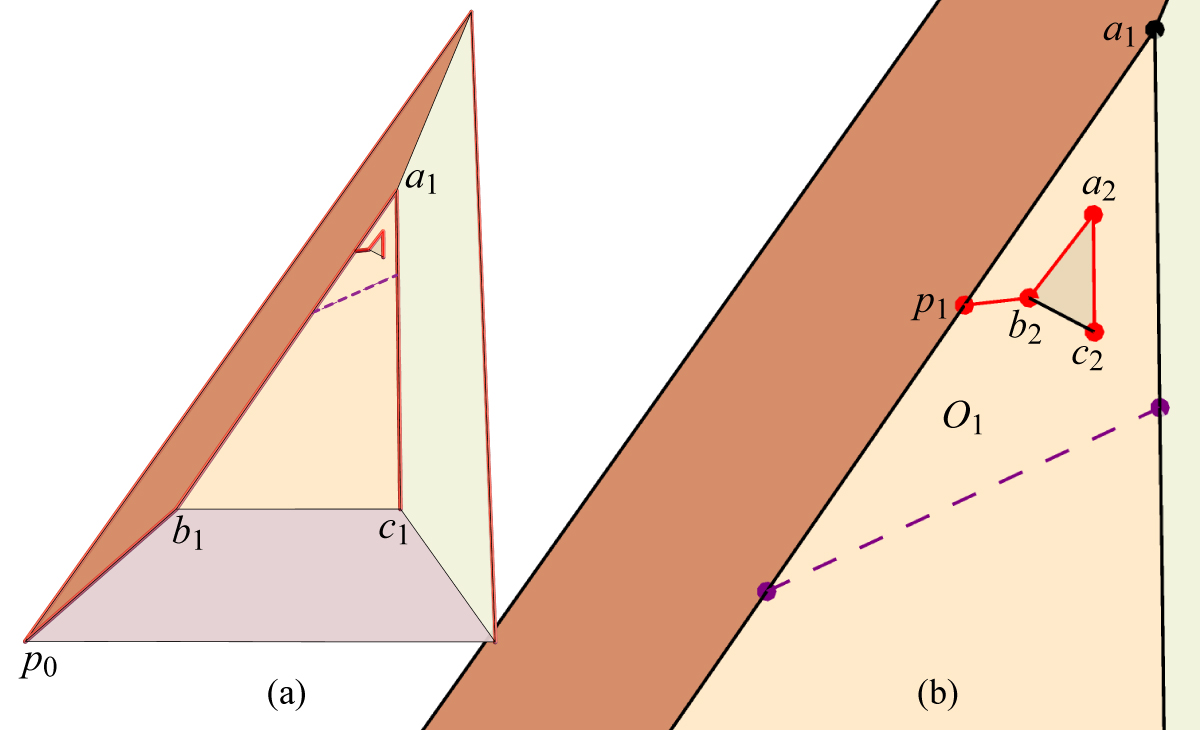}
\caption{Polyhedron $P_2$.
(a)~Cut tree (red).
(b)~Enlarged view of $\triangle_2$ floating over overlap region. 
The convex hull is shown ahead in Fig.~\protect\figref{Vertex_p}.
\anna{The dashed line is the extent of the first\ff{-}level overlap, $O_1$.}
}
\figlab{Gadget_2nd_3D}
\end{figure}

\Anna{After describing $P_2$ as above, please describe the cut and note that it is a tree not a path.}
\jor{Note that the cuts to produce $P_2$ form a tree, not a path.}

The unfolding is shown in Fig.~\figref{Gadget_2nd_2D_all}.
Because the modifications to $P_1$ were small, $P_2$ looks very close to the unfolding of
$P_1$ (Fig.~\figref{Gadget_1st_2D3D}(a)).
To see the difference, focus on the enlarged view in Fig.~\figref{Gadget_2nd_2D_enlarged}.
The cut path $p_1,b_2,a_2,c_2$ creates an overlap in a neighborhood of $a_2$.
Within that overlap region, we can embed a smaller overlap gadget.

\Anna{I suggest devoting a sentence to describing $P_i$.}
\jor{$P_i$ consists of stacked and nested triangles $\triangle_i$: stacked in parallel planes at height $z_i$, nested within each level's overlap region $O_i$.}
This nesting can continue to achieve any desired thickness of overlap $t$.

\begin{figure}[!h]
\centering
\includegraphics[width=0.8\columnwidth]{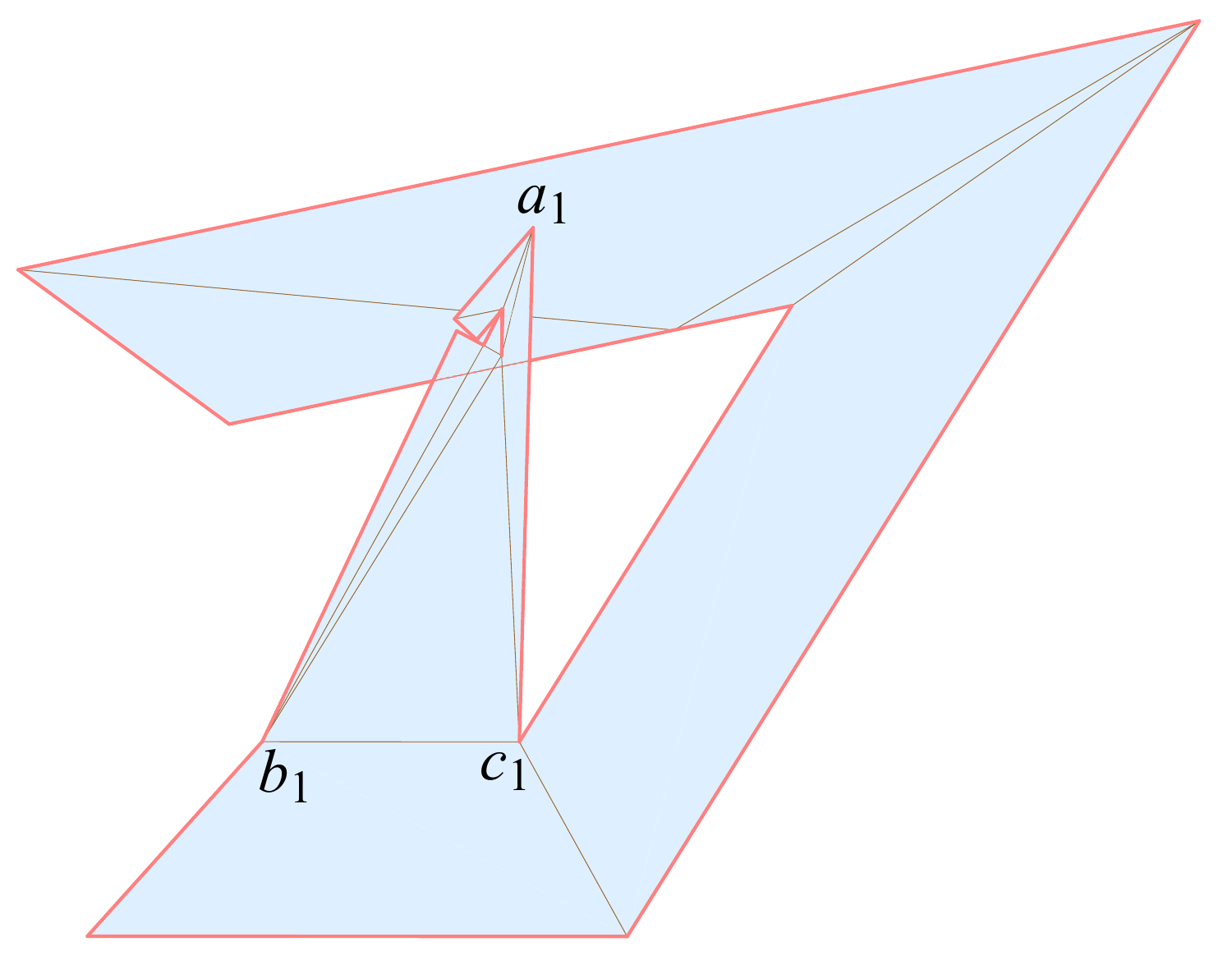}
\caption{Unfolding of $P_2$
\anna{from Fig.~\protect\figref{Gadget_2nd_3D}.}
\Anna{The red line should be dashed (or hidden) from $o$ to $o'$ (BTW those labels are never used in the text)}
\JOR{Good catch. Fixed hidden lines in oerlap.}
}
\figlab{Gadget_2nd_2D_all}
\end{figure}

\begin{figure}[htbp]
\centering
\includegraphics[width=0.75\columnwidth]{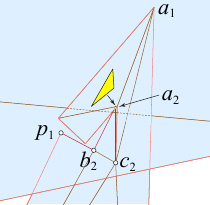}
\caption{Enlarged view of overlap area in Fig.~\protect\figref{Gadget_2nd_2D_all}.
\Anna{The edges going from far left to far right should be dashed as they go under the overlap (right now only the top one is dashed.)}
}
\figlab{Gadget_2nd_2D_enlarged}
\end{figure}

\subsection{Nesting Analyzed}
Now we argue that this plan of nesting overlap gadgets is sound. \Anna{It might help to clarify what exactly needs justification. The analysis in step 4 is clear and we need it.  Some of the other points seem more like expanding on the construction of $P_2$.  I guess we need that the construction of $P_i$ produces a valid convex polyhedron and that the cut unfolds as illustrated in the figures.}\JOR{Not sure how to address this. To make the gadget work, $\omega$ must be small and the angle at $a_2$ $<60^\circ$. $\triangle_i$ must fit under the dome formed by extension of faces. And the cut path has to follow edges. 
Looking at my Mathematica calcs, I had to hack-adjust the four vertices $p_1,a_2,b_2,c_2$ to get it to work.
Here's a try:}
\jor{Focusing on $i=2$, we need $\omega$ and the angle at $a_2$ of $\triangle_2$ to be such that the gadget does indeed create overlap.
And we need $\triangle_2$ placed so that $P_2$ includes the cut path that unfolds as illustrated in Fig.~\figref{Gadget_2nd_2D_all}.
More precisely:}
\begin{enumerate}[(1)]
\item As mentioned, $\omega$ can be selected arbitrarily small by reducing the
height $z_2$ of $\triangle_2$ 
over $\triangle_1$. 
In the example, $\omega \approx 1.5^\circ$ at $c_2$.
\item \Anna{Please clarify.  Is this intended to state a requirement? How is it met?  As I understand it, $a_1 c_1$ is an edge of the polyhedron (so nonflat dihedral), and $a_1 b_1$ is no longer an edge -- rather it is broken into $a_1 p_1$ and $p_1 b_1$, both nonflat.}
\jor{The edges $a_1 c_1$, $a_1 p_1$, and $p_1 b_1$ of $P_1$ are all nonflat.
Extending their incident faces above the plane of $\triangle_1$ creates a convex ``canopy'' under which $\triangle_2$ can be placed. This placement, together with choosing $\omega$ (see (1) above), determines $z_2$.
}

\fade{
There must be some nonflat dihedral angle along edges $a_1 b_1$ and $a_1 c_1$,
so that there is room above $\triangle_1$ and below the extension of the associated faces
for the construction to remain convex.
}
\item 
After altering $P_1$ by installing $\triangle_2$ 
parallel to and above $\triangle_1$ in the overlap region $O_1$, 
and connecting $\triangle_2$ to $\triangle_1$
by the edge $p_1 b_2$, the new additions need to be ``retriangulated.''
One can view this as taking the convex hull of $\triangle_1$ and the four added vertices.
\Anna{I see.  The quadrilateral $a_1 p_1 b_2 a_2$ is not flat.  But the other quads (on $a_2 c_2$ and $b_2 c_2$) can be flat if $\triangle_2$'s sides are parallel to $\triangle_1$'s sides}
\Anna{This was said earlier as part of the construction:} What we need to achieve for an edge unfolding is that the cut path $(p_1,b_2,a_2,c_2)$ follows edges of the polyhedron $P_2$.

For the generic situation, $\triangle_{i+1}$ sits over $\triangle_i$, and $p_i$ lies
just outside $a_i b_i$. Then the convex hull always includes $p_i b_{i+1}$,
as shown in Fig.~\figref{Vertex_p}, and of course the edges of $\triangle_{i+1}$ are included.
Because the structure of $\triangle_{i+1}$ over $\triangle_i$ is just a scaled version of
$\triangle_{2}$ over $\triangle_1$, the 
cut path indeed follows edges of the polyhedron $P_{i+1}$.
\erik{Changed $P$ index to $i+1$; correct?}
\JOR{I *think* so, and I'm afraid to change all indexing.}

\begin{figure}[htbp]
\centering
\includegraphics[width=0.60\columnwidth]{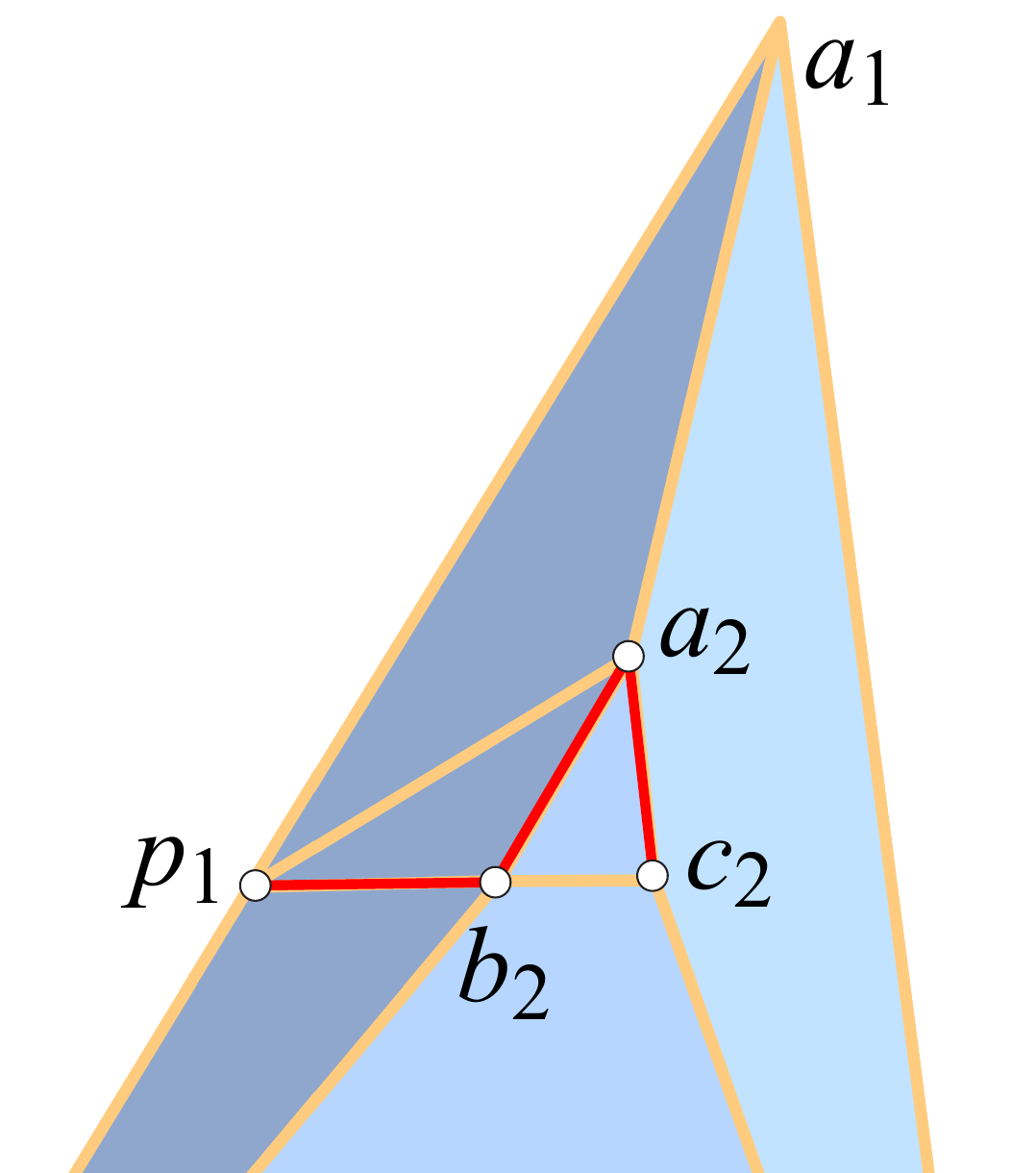}
\caption{The 3D convex hull shows that the branch of the cut path $(p_1,b_2,a_2,c_2)$
follows polyhedron edges.}
\figlab{Vertex_p}
\end{figure}

\item The claim that $n = O(t)$ follows because each nesting adds four vertices.
More precisely, $n = 6 + 4 (t-2)$.
\erik{This formula only handles $t \geq 2$.}
$P_1$ (Fig.~\figref{Gadget_1st_2D3D}) has $n=6$ vertices and achieves $t=2$.
$P_2$ (Fig.~\figref{Gadget_2nd_3D}) has $n=10$ vertices and achieves $t=3$.
\end{enumerate}

\noindent
This completes the proof of Theorem~\thmref{PolyhedronOverlap}.

\JOR{I think there might be now some redundancy in the descriptions above. But better redundant than missing points.}

\section{Conclusion}
%
A natural next step is to determine the largest 
overlap thickness achievable for specific polyhedra.
The Platonic solids do not have edge unfoldings that overlap~\cite{hs-eupsn-11},  
but in a forthcoming paper
we analyze the thickness of general unfoldings of the Platonic solids and other polyhedra.  
\jor{For example, in related work}
Shiota and Saitoh~\cite{shiota2024overlapping} found that
the truncated dodecahedron has an edge unfolding with
overlap thickness $2$. 
\Anna{Do we want to put this here?  or is it better in our forthcoming paper.}\JOR{I would include here.}
Is that the largest thickness achievable
for that polyhedron?





\newpage
\small
\bibliographystyle{abbrv}
\bibliography{refs} 



\end{document}